\documentclass[journal]{IEEEtran}
\usepackage{cite}
\usepackage{amsmath,amssymb,amsfonts}
\usepackage{algorithmic}
\usepackage{graphicx}
\usepackage{amssymb,comment}
\usepackage{algorithmic}
\usepackage{algorithm}
\usepackage{caption} 
\usepackage{subfig}
\usepackage{subcaption}
\usepackage{scrextend}
\usepackage{verbatim}
\usepackage{color}

\newtheorem{theorem}{\bf Theorem}

\newcommand{\ic}[1]{\in\mathcal{#1}}

\usepackage{setspace}
\begingroup
\allowdisplaybreaks

\begin{document}
\title{Joint Beamforming and Antenna Position Optimization for Movable Antenna-Assisted Spectrum Sharing}

\author{Xin~Wei,
        Weidong~Mei,~\IEEEmembership{Member,~IEEE,}
        Dong~Wang,~\IEEEmembership{Student Member,~IEEE,}
        Boyu Ning,~\IEEEmembership{Member,~IEEE,}
        and~Zhi~Chen,~\IEEEmembership{Senior Member,~IEEE}
\thanks{The authors are with the National Key Laboratory of Wireless Communications, University of Electronic Science and Technology of China, Chengdu, China (e-mail: xinwei@std.uestc.edu.cn, wmei@uestc.edu.cn, DongwangUESTC@outlook.com, boydning@outlook.com, chenzhi@uestc.edu.cn).}}

\maketitle
\begin{abstract}
Fluid antennas (FAs) and movable antennas (MAs) have drawn increasing attention in wireless communications recently due to their ability to create favorable channel conditions via local antenna movement within a confined region. In this letter, we advance their application for cognitive radio to facilitate efficient spectrum sharing between primary and secondary communication systems. In particular, we aim to jointly optimize the transmit beamforming and MA positions at a secondary transmitter (ST) to maximize the received signal power at a secondary receiver (SR) subject to the constraints on its imposed co-channel interference power with multiple primary receivers (PRs). However, such an optimization problem is difficult to be optimally solved due to the highly nonlinear functions of the received signal/interference power at the SR/all PRs in terms of the MA positions. To drive useful insights, we first perform theoretical analyses to unveil MAs' capability to achieve maximum-ratio transmission with the SR and effective interference mitigation for all PRs at the same time. To solve the MA position optimization problem, we propose an alternating optimization (AO) algorithm to obtain a high-quality suboptimal solution. Numerical results demonstrate that our proposed algorithms can significantly outperform the conventional fixed-position antennas (FPAs) and other baseline schemes.
\end{abstract}
\IEEEpeerreviewmaketitle

\section{Introduction}
The cognitive radio (CR) technique enables secondary users to utilize spectrum resources originally allocated to primary users (PUs), without significantly affecting the quality of service experienced by PUs. This approach has gained widespread adoption in wireless communications, primarily due to the escalating challenges posed by constrained spectrum resources. \cite{R_Zhang_Dynamic}.

In this letter, we advance the use of the emerging fluid antenna (FA)/movable antenna (MA) technology for CR. As compared to the conventional fixed-position antennas (FPAs), the FA/MA technology allows for flexible antenna movement within a confined region at the transmitter/receiver to create favorable channel conditions, thereby greatly enhancing the wireless communication performance \cite{L_Zhu_Movable,F_Ghadi_Fluid}. Inspired by the promising benefits of the FAs/MAs, some existing works have delved into the FAs'/MAs' position optimization under various system setups. Particularly, the authors in \cite{W_Ma_Multi_beam} and \cite{L_Zhu_MA_Null_Steering} investigated the capability of a linear MA array to achieve flexible beam forming under the line-of-sight channel conditions. In addition, the FA/MA position optimization has also been studied in other scenarios, including multiple-input multiple-output (MIMO) system \cite{W_Ma_MA_MIMO}, physical-layer security \cite{B_Tang_Fluid}, non-orthogonal multiple access \cite{W_New_Fluid}, over-the-air computation \cite{D_Zhang_Air}, etc. Instead of optimizing the MA positions in a continuous space as above, the authors in \cite{W_Mei_Graph} proposed to discretize the transmit region into multiple sampling points and introduced a graph-based method to select an optimal subset of the sampling points to maximize the communication rate. However, to the best of our knowledge, there is no existing work focusing on the MA position optimization in CR.

\begin{figure}[!t]
\centering
\captionsetup{justification=raggedright,singlelinecheck=false}
\centerline{\includegraphics[width=0.5\textwidth]{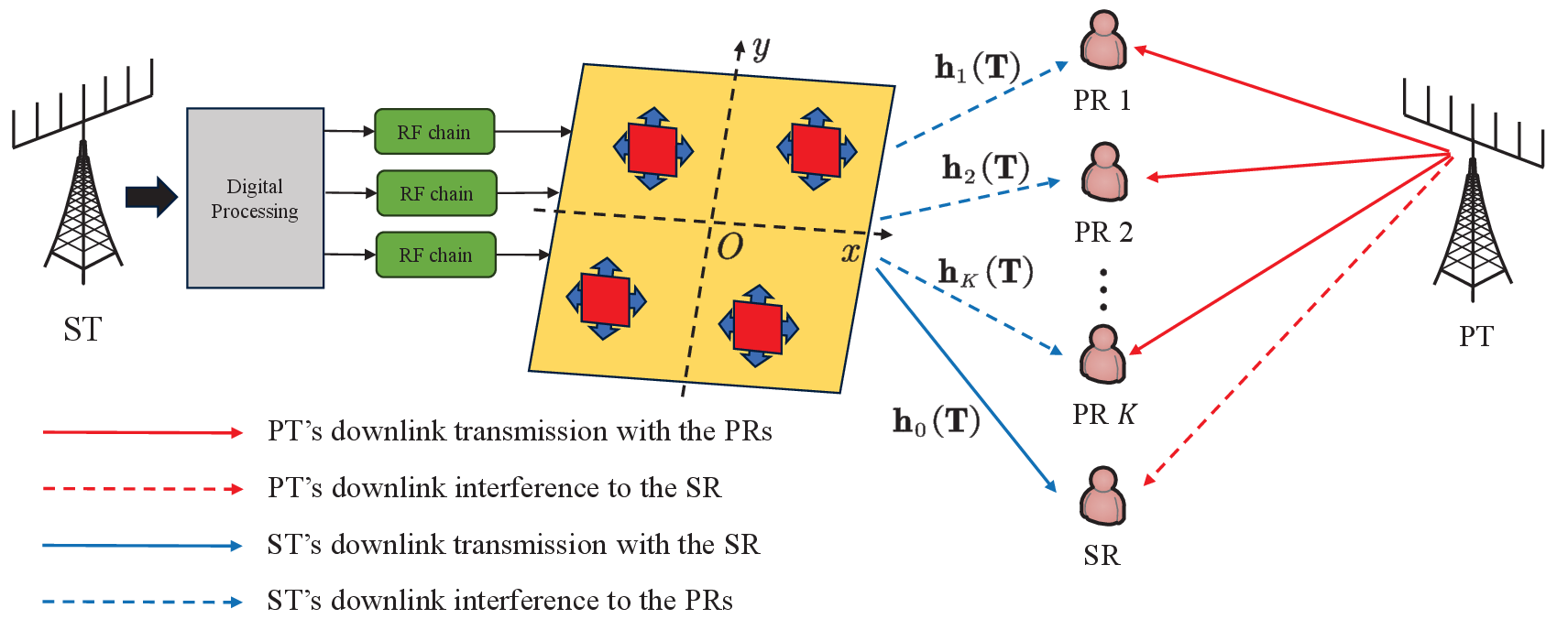}}
\caption{MA-enhanced CR system.}
\label{Fig_SysModel}
\vspace{-23pt}
\end{figure}

To fill in this gap, in this letter, we investigate the MA position optimization in CR system as shown in Fig. 1, where an second transmitter (ST) equipped with multiple MAs communicates with an second receiver (SR) in the presence of multiple co-channel primary receivers (PRs). To ensure spectrum-efficient coexistence between the secondary and the primary communications, we aim to jointly optimize the transmit beamforming and MA positions at the ST to maximize the received signal power at the SR subject to the constraints on its imposed co-channel interference to the PRs. However, such an optimization problem is difficult to be optimally solved as the received signal/interference power at the SR/each PR appears to be a highly nonlinear function of the MA positions. To gain essential insights, we first conduct theoretical analyses to show the capability of MAs to achieve maximum-ratio transmission (MRT) with the SR and effective interference mitigation for all PRs at the same time under certain conditions. Note that our derived results extend those in \cite{L_Zhu_MA_Null_Steering} for beam nulling to a more general case with multi-path channels and a given interference power limit. Then, we propose an alternating optimization (AO) algorithm to solve the MA position optimization problem by combining the successive convex approximation (SCA) and the discrete sampling approaches, which are easier to apply than the existing continuous location update algorithms \cite{W_Ma_MA_MIMO,Z_Xiao_Multiuser}. Numerical results demonstrate that our proposed algorithm can achieve much better performance as compared to the conventional FPAs and other baseline schemes.

\section{System Model and Problem Formulation}
As shown in Fig. 1, we consider an MA-assisted CR system, where a primary transmitter (PT) communicates with $K\ge 1$ PRs simultaneously in the presence of the transmission from an ST to an SR that operates over the same frequency band\footnote[1]{Note that the results in this paper are also applicable to the case of multiple SRs with an orthogonal time allocation for them.}. Assume that the SR and PRs are all equipped with a single FPA, considering their much smaller antenna regions compared to the ST (e.g., a base station), while the PT is equipped with multiple FPAs with its transmit beamforming already optimized to serve the $K$ PRs\footnote[2]{In this letter, the primary communication system can be viewed as an existing wireless network that applies the conventional FPA architecture.}. As such, it can be equivalently treated as being equipped with a single FPA. Let $\mathcal{K}=\{1,2,\cdots,K\}$ denote the set of all PRs. To mitigate the co-channel interference from the ST to the PRs in $\cal K$, we consider that the ST is equipped with $N$ MAs that can be flexibly moved within a 2D transmit region, which is a square region of size $A \times A$ and denoted as ${\cal C}_t$. Without loss of generality, we assume that $\mathcal{C}_t$ is parallel to the $x$-$y$ plane, as shown in Fig. 1. Denote by $\mathbf{t}_n=[x_n,y_n]^T\in\mathcal{C}_t, n \in {\cal N}\triangleq\{1,2,\cdots,N\}$, the position of the $n$-th transmit MA in $\mathcal{C}_t$, given a reference origin point $\mathbf{t}_0= [0, 0]^T$.

We assume that all the involved channels are quasi-static, such that the switching delay for antenna movement is tolerable compared with the channel coherence time. To characterize the channels from the ST to the SR and PRs, we adopt the array-response based channel model as in \cite{L_Zhu_Modeling}. In particular, by stacking the $N$ MAs' positions at the ST, we define its antenna position vector (APV) as $\mathbf{T}=[\mathbf{t}_1,\mathbf{t}_2,\cdots,\mathbf{t}_N]\in\mathbb{C}^{2\times N}$ and denote its downlink channels with the SR and PR $k$ by $\mathbf{h}^H_0(\mathbf{T}) \in \mathbb{C}^{1\times N}$ and $\mathbf{h}^H_k(\mathbf{T}) \in \mathbb{C}^{1\times N}$, $\forall k \in \cal K$, respectively. Let $L_k$ be the number of transmit paths from the ST to PR $k$, and $\theta_{k,p}$/$\phi_{k,p}$ be the elevation/azimuth angle of departure (AoD) for the $p$-th ($p = 1,2,\cdots,L_k$) transmit path. Then, the array response of the $p$-th transmit path from the ST to PR $k$ can be written as
\begin{equation}\label{eqn_ArrayResp}
\begin{aligned}
    \mathbf{a}\left( \mathbf{T}, \theta_{k,p},\phi_{k,p} \right) = [& e^{j\rho _{k,p}\left( \mathbf{t}_1 \right)},e^{j\rho _{k,p}\left( \mathbf{t}_2 \right)},
    \cdots ,e^{j\rho _{k,p}\left( \mathbf{t}_N \right)} ] ,
\end{aligned}
\end{equation}
where $\lambda$ denotes the wavelength, and $\rho _{k,p}\left( \boldsymbol{\mathrm{t}}_n \right) =\frac{2\pi}{\lambda}(x_n\sin \theta _{k,p}\cos \phi _{k,p}+y_n\cos \theta _{k,p})$ denotes the phase difference between the $n$-th MA and $\boldsymbol{\mathrm{t}}_0$. As such, the channel vector from the ST to PR $k$ can be expressed as
\begin{equation}\label{eqn_Channel}
    \mathbf{h}_k^H\left( \mathbf{T} \right) =\sum_{p=1}^{L_{k}}{\beta_{k,p}\mathbf{a}^H\left( \mathbf{T}, \theta _{k,p},\phi _{k,p} \right) }, k\ic{K},
\end{equation}
where $\beta_{k,p}$ denotes the path gain of the $p$-th transmit path from the ST to PR $k$. Similarly, the channel from the ST to the SR, i.e., $\boldsymbol{\mathrm{h}}_0(\mathbf{T})$, can be defined as in (\ref{eqn_Channel}) by replacing the subscript ``$k$'' therein with ``$0$''. It is worth noting that, unlike the statistical channel models studied in \cite{F_Ghadi_Fluid,B_Tang_Fluid,W_New_Fluid,D_Zhang_Air} for rich-scattering environments, we consider a sparse environment with a limited number of scatterers. In this model, any two nearby antenna positions in ${\cal C}_t$ tend to achieve more similar channels with the SR/each PR due to a more similar geometric relationship with the scatterers compared to two distant antenna positions.

Let $\mathbf{w}\in\mathbb{C}^{N\times 1}$ denote the transmit beamforming vector of the ST, with $||\mathbf{w}||^2_2 \le P_{\max}$ and $P_{\max}$ being its maximum transmit power. Then, the received signal-to-noise ratio (SNR) at the SR is given by
\begin{equation}\label{eqn_SR_SINR}
    \gamma _{\mathrm{SR}}=\frac{\left|\mathbf{h}_0^H\left( \mathbf{T}\right) \mathbf{w} \right| ^2}{\sigma ^2},
\end{equation}
where $\sigma^2$ comprises the total power of background noise and the PT's interference at SR (both assumed to be independently circle symmetric complex Gaussian (CSCG) distributed).

Due to the spectrum sharing between the primary and the secondary transmission, the ST may impose severe co-channel interference to the PRs. To ensure the communication performance of the PRs, we apply the celebrated interference temperature (IT) technique in CR, such that the received interference power at each PR, i.e.,  $\left|\mathbf{h}_k^H\left( \mathbf{T}\right) \mathbf{w} \right| ^2$, does not exceed a prescribed threshold, i.e.,
\begin{equation}\label{eqn_Intf_Cons}
    \left|\mathbf{h}_k^H\left( \mathbf{T}\right) \mathbf{w} \right| ^2\le \Gamma , \forall k \in \cal K, 
\end{equation}
where $\Gamma$ denotes the prescribed limit of the interference power.

In this letter, we aim to maximize the received SNR at the SR, i.e., $\gamma_{\mathrm{SR}}$ in (\ref{eqn_SR_SINR}), subject to the IT constraints in (\ref{eqn_Intf_Cons}), by jointly optimizing the ST's transmit beamforming vector $\boldsymbol{\mathrm{w}}$ and the APV $\mathbf{T}$. Thus, the problem is formulated as
\begin{subequations}\label{eqn_OptPrblm_P1}
\begin{align}
{\text{(P1)}}\quad &\underset{\boldsymbol{\mathrm{w}},\mathbf{T}} {\max}\quad \gamma_{\mathrm{SR}} \nonumber
\\
\mathrm{s.t.}\quad & \boldsymbol{\mathrm{t}}_n \in \mathcal{C}_t, n \in \mathcal{N}, \label{eqn_MA_Region_Cons}
\\
& ||\boldsymbol{\mathrm{t}}_k - \boldsymbol{\mathrm{t}}_m||_2 \ge D_{\min}, k,m\in\mathcal{N}, k\ne m \label{eqn_MA_Coordinate_Cons}
\\
& ||\boldsymbol{\mathrm{w}}||^2_2 \le P_{\max}, \label{eqn_BeamFormer_Cons}
\\
& \left|\mathbf{h}_k^H\left( \mathbf{T}\right) \mathbf{w} \right| ^2\le \Gamma , \forall k \in \cal K,
\end{align}
\end{subequations}
where $D_{\min}$ denotes the minimum spacing between any two MAs to avoid the mutual coupling. It is worth noting that to investigate the fundamental limit of the proposed MA-assisted CR, we assume that all required channel state information are available by applying some channel estimation techniques dedicated to MAs/FAs, e.g., \cite{W_Ma_Compressed} and \cite{K_Wong_Virtual}.

However, it is noted that (P1) is difficult to be optimally solved due to the intricate coupling between the transmit beamforming $\mathbf{w}$ and the antenna positions $\mathbf{T}$. In the next section, we first perform some theoretical analyses to show the efficacy of using MAs for interference mitigation in CR and draw useful insights, which also provide optimal solutions to (P1) in some special cases. In Section IV, we will solve (P1) in the general case.

\section{Theoretical Analysis}
In this section, we conduct theoretical analyses to reveal the capability of MAs to achieve MRT and effective interference mitigation at the same time. To facilitate our analysis, we assume in this section that the size of the transmit region $\mathcal{C}_t$ (i.e., $A$) is sufficiently large. Moreover, we assume that the AoD for each transmit path associated with any PR $k$ is different from that for any transmit path associated with the SR, i.e., $\theta_{k,p}\ne\theta_{0,q}$, $\phi_{k,p}\ne\phi_{0,q}$, $q=1,2,\cdots,L_0,p=1,2,\cdots,L_k$, $\forall k\in\mathcal{K}$.

Specifically, if the ST applies the MRT, its transmit beamforming vector for any given APV $\mathbf{T}$, is written as
\begin{equation}\label{eqn_MRT_BF}
    \mathbf{w}_{\text{MRT}}(\mathbf{T})=\frac{\sqrt{P_{\max}}}{||\mathbf{h}_0(\mathbf{T})||_2}\mathbf{h}_0(\mathbf{T}),
\end{equation}
and the resulting co-channel interference power at PR $k$ is given by
\begin{equation}\label{eqn_InfPow}
    \begin{aligned}
        &P_k(\mathbf{T})=\frac{P_{\max}}{||\mathbf{h}_0(\mathbf{T})||_2^2}\left|\sum_{p=1}^{L_k}{\mathbf{a}^H\left( \mathbf{T}, \theta_{k,p},\phi_{k,p} \right)\mathbf{h}_0(\mathbf{T})}\right|^2\\
        &=\frac{P_{\max}}{||\mathbf{h}_0(\mathbf{T})||_2^2}\left|\sum_{p=1}^{L_k}\sum_{q=1}^{L_0}{\beta_{k,p}^*\beta_{0,q}\sum_{n=1}^{N}{e^{j\frac{2\pi}{\lambda}(a_{k,p}^qx_n+b_{k,p}^qy_n)}}}\right|^2,\forall k,
    \end{aligned}
\end{equation}
where
\begin{equation}
    \begin{aligned}
        a_{k,p}^q&\triangleq\sin \theta _{0,q}\cos \phi _{0,q}-\sin \theta _{k,p}\cos \phi _{k,p},\,\,\forall k,p,q\\
        b_{k,p}^q&\triangleq\cos \theta _{0,q}-\cos \theta _{k,p},\,\,\forall k,p,q.\nonumber
    \end{aligned}
\end{equation}
Next, we show that there exists an APV solution $\mathbf{T}^{\star}$, such that $P_k(\mathbf{T}^{\star})\le\Gamma$, $\forall k$, i.e., the IT constraints in (\ref{eqn_Intf_Cons}) are met, under certain conditions. To this end, we first consider a simplified case with $N=2$ and present the following theorem.

\begin{theorem}
    When $N=2$, there must exist an APV solution to (P1), denoted as $\mathbf{T}^{\star}$, such that $P_k(\mathbf{T}^{\star})\le\Gamma$, $\forall k$.
\end{theorem}
\begin{IEEEproof}
    To construct $\mathbf{T}^\star$, we consider placing all MAs along the $y$-axis with an equal spacing of $d_y$. As such, the position of the $n$-th MA is given by $\mathbf{t}_n=[0,(n-1)d_y]^T$. By setting $x_n=0,y_n=(n-1)d_y$ and $N=2$ in (\ref{eqn_InfPow}), the co-channel interference power can be simplified as
    \begin{equation}\label{eqn_Th1_Eq1}
        P_k(\mathbf{T})=\frac{P_{\max}}{||\mathbf{h}_0(\mathbf{T})||_2^2}\left|\sum_{p=1}^{L_k}\sum_{q=1}^{L_0}{\beta_{k,p}^*\beta_{0,q}\left(1+e^{j\frac{2\pi}{\lambda}d_yb_{k,p}^q}\right)}\right|^2,\forall k.
    \end{equation}
    To proceed, let $H_{\min}$ denote the minimum value of $||\mathbf{h}_0(\mathbf{T})||_2^2$ among all feasible APV solutions to (P1), i.e., $H_{\min}=\min_{\mathbf{T}}||\mathbf{h}_0(\mathbf{T})||_2^2$. Thanks to the use of MRT, it must hold that $H_{\min} >0$. Let $\gamma\triangleq\frac{P_{\max}}{H_{\min}}>0$. Then, the co-channel interference power $P_k(\mathbf{T})$ in (\ref{eqn_Th1_Eq1}) can be upper-bounded by
    \begin{equation}\label{eqn_InfPower}
        \begin{aligned}
            P_k(\mathbf{T})&\le \gamma \left|\sum_{p=1}^{L_k}\sum_{q=1}^{L_0}{\beta_{k,p}^*\beta_{0,q}\left(1+e^{j\frac{2\pi}{\lambda}d_yb_{k,p}^q}\right)}\right|^2\\
            &\le2\gamma\left|\sum_{p=1}^{L_k}\sum_{q=1}^{L_0}|\beta_{k,p}||\beta_{0,q}|\sqrt{1+\cos\frac{2\pi}{\lambda}d_yb_{k,p}^q}\right|^2,\forall k,
        \end{aligned}
    \end{equation}
    where the second inequality holds if and only if all of the $L_kL_0$ terms are co-phase. Next, we introduce the following lemma.
    
    \textbf{Lemma 1} \cite{A_Leshem_InfChannel}: Let $g(b_{k,p}^q,d_y)=\frac{1}{2}(1+\cos\frac{2\pi}{\lambda}d_yb_{k,p}^q)$, $\forall k,p,q$. If all $b_{k,p}^q$'s are independent over the set of rational numbers, then for any $\delta>0$, there must exist a positive integer spacing $d_y$ such that
    \begin{equation}
        g(b_{k,p}^q,d_y) < \delta,\,\,\forall k,p,q.
    \end{equation}

    Note that all $b_{k,p}^q$'s generally take an irrational value in practice considering the random locations of the environment scatterers. As such, they are independent over the set of rational numbers with the probability of one. Based on Lemma 1, for any given $\delta > 0$ there must exist an integer $d_y$ such that $\frac{1}{2}(1+\cos\frac{2\pi}{\lambda}d_yb_{k,p}^q) < \delta, \forall k,p,q$. By substituting these inequalities into (\ref{eqn_InfPower}), we have
    \begin{equation}\label{eqn_Th1_Eq3}
        P_k(\mathbf{T})\le 4\gamma\delta\left|\sum_{p=1}^{L_k}\sum_{q=1}^{L_0}|\beta_{k,p}||\beta_{0,q}|\right|^2, \forall k.
    \end{equation}
    If the right-hand side of (\ref{eqn_Th1_Eq3}) is smaller than $\Gamma$ for all $k\ic{K}$, the IT constraints in (\ref{eqn_Intf_Cons}) will be met. This can be realized by setting
    \begin{equation}
        \delta = \underset{k\ic{K}}{\min}\frac{\Gamma}{4\gamma\left|\sum_{p=1}^{L_k}\sum_{q=1}^{L_0}|\beta_{k,p}||\beta_{0,q}|\right|^2},
    \end{equation}
    and apply Lemma 1 accordingly. Notably, as $d_y$ is a positive integer, it should be much greater than the wavelength-level antenna spacing in general, especially for future wireless communication systems migrating to higher operating frequency bands with shorter wavelength. Hence, the inter-MA spacing constraints in (5a) should be satisfied. This thus completes the proof.
\end{IEEEproof}

Theorem 1 indicates that for a sufficiently large transmit region, the MRT can be applied to enhance the rate performance of the SR while suppressing the co-channel interference to the PRs. It follows from the above that antenna position optimization offers more degrees of freedom to achieve higher flexibility of beamforming as compared to the conventional FPAs.

However, Theorem 1 only holds for $N=2$. Next, we show that in the special case of $L_0=1$, we may achieve simultaneous MRT and interference nulling for a general $N$. In this case, the array response in (\ref{eqn_Channel}) reduces to the following vector w.r.t. $(\theta_{0,1},\phi_{0,1})$, i.e.,
\begin{equation}\label{eqn_SV_SR}
\mathbf{a}_0\left( \mathbf{T}, \theta _{0,1}, \phi _{0,1} \right) =[ e^{j\rho _{0,1}\left( \boldsymbol{\mathrm{t}}_1 \right)},e^{j\rho _{0,1}\left( \boldsymbol{\mathrm{t}}_2 \right)},\cdots ,e^{j\rho _{0,1}\left( \boldsymbol{\mathrm{t}}_N \right)} ] ^T,
\end{equation}
where $\rho_{0,1}\left( \mathbf{t}_n \right)=\frac{2\pi}{\lambda}(x_n\sin \theta _{0,1}\cos \phi _{0,1}+y_n\cos \theta _{0,1})$.

If the MRT is still used for the ST's transmit beamforming, it can be shown that the resulting beam gain from the ST to the $p$-th transmit path of PR $k$ is given by
\begin{equation}
    G(\mathbf{T},\theta_{k,p},\phi_{k,p}) = \left|\sum_{n=1}^N{\exp\left[j\frac{2\pi}{\lambda}\left(x_na_{k,p}^1+y_nb_{k,p}^1\right)\right]}\right|^2, \forall k,p.
\end{equation}
Denote by $L_{\text{tot}}=\sum_{k=1}^K{L_k}$ the total number of the transmit paths from the ST to all PRs \footnote[1]{Note that the actual total number of paths may be smaller than $L$ since some PRs and the SR may share the common transmit paths. For convenience, we consider the worst case that all transmit paths for them are different.}. Next, we show that interference nulling over all paths from the ST to the PRs can be achieved under certain conditions on $L_{\text{tot}}$.

\begin{theorem}
Denote the prime factorization of $N$ as $N=\prod_{i=1}^{I_N}f_i$, where $I_N$ represent the total number of prime factors of $N$ and they are sorted in a non-decreasing order as $2\leq f_1\leq f_2\leq\cdots\leq f_{I_N}$. Then, an APV $\mathbf{T}^{*}$ satisfying $G(\mathbf{T}^{*},\theta_{k,p},\phi_{k,p})=0$ and constraint (5b) always exists for if $L_{\text{tot}}\leq I_N$.
\end{theorem}
\begin{IEEEproof}
    Theorem 2 is an extension of Proposition 1 in \cite{L_Zhu_MA_Null_Steering} from 1D transmit region to 2D. It can be similarly proved as in \cite{L_Zhu_MA_Null_Steering} by fixing the vertical dimension of all MAs (i.e., $y_n,n\ic{N}$) as any given value. Then, the horizontal dimension of all MAs, i.e., $x_n, n\ic{N}$, can be constructed in a successive manner. The details can be found in \cite{L_Zhu_MA_Null_Steering} and omitted in this letter for brevity.
\end{IEEEproof}

Theorem 2 indicates that for any given number of MAs, the desired flexible beamforming can be achieved if the number of transmit paths is sufficiently small. Since $I_N \ge 1$, it must be achieved in the case of a common single transmit path from the ST to all PRs.

\section{Proposed Solutions to (P1)}
In this section, we aim to solve problem (P1) in the general case with a finite length of $A$. To this end, we propose an AO algorithm to decompose (P1) into two subproblems and solve them alternately. 

\subsection{Optimizing $\mathbf{w}$ for Given $\mathbf{T}$}
First, we aim to optimize the beamforming vector $\mathbf{w}$ with given MAs' positions $\mathbf{T}$. In this case, as each $\mathbf{h}_k(\mathbf{T}), k \in {\cal K}$, is fixed, problem (P1) can be simplified as
\begin{subequations}\label{eqn_OptPrblm_P2}
\begin{align}
{\text{(P2)}}\quad &\underset{\boldsymbol{\mathrm{w}}} {\max}\quad \mathbf{w}^H \mathbf{H}_0(\mathbf{T})\mathbf{w} \nonumber
\\
\mathrm{s.t.}\quad & \mathbf{w}^H \mathbf{w} \le P_{\max}, \label{eqn_BeamFormer_Cons_P2}
\\
& \mathbf{w}^H \mathbf{H}_k(\mathbf{T}) \mathbf{w} \le \Gamma, \forall k \in \cal K, \label{eqn_IntCons_P2}
\end{align}
\end{subequations}
where $\mathbf{H}_k(\mathbf{T})\triangleq\mathbf{h}_k(\mathbf{T}) \mathbf{h}_k^H(\mathbf{T})$ for $k\ic{K} \cup \{0\},$ and the noise power $\sigma^2$ in the objective function is omitted. However, problem (P2) is still a non-convex optimization problem as its objective function is convex (instead of concave) in $\mathbf{w}$. To tackle this challenge, we apply the SCA technique to transform (P2) into a series of more tractable approximated convex subproblems. Specifically, with a given local point $\mathbf{w}_i$, the objective function of (P1) can be lower-bounded by its first-order Taylor expansion at $\mathbf{w}_i$, which is given by
\begin{align}
\mathbf{w}^H \mathbf{H}_0(\mathbf{T}) \mathbf{w}&\ge \mathbf{w}_i^H\mathbf{H}_{0}(\mathbf{T})\mathbf{w}_i +2\mathrm{Re}\left\{ \mathbf{w}_i^H \mathbf{H}_{0}(\mathbf{T})\left( \mathbf{w}-\mathbf{w}_i \right) \right\} 
\nonumber\\
&=2\underset{\triangleq f\left( \mathbf{w} \right)}{\underbrace{\mathrm{Re}\left\{  \mathbf{w}_i^H \mathbf{H}_{0}(\mathbf{T}) \mathbf{w} \right\} }}-\underset{\text{constant}}{\underbrace{\mathbf{w}_i^H \mathbf{H}_{0}(\mathbf{T}) \mathbf{w}_i}}.\label{eqn_ObjFun_SCA_P2}
\end{align}
By omitting the constant term in (\ref{eqn_ObjFun_SCA_P2}), problem (P2) in the $i$-th SCA iteration is given by
\begin{equation}\label{eqn_OptPrblm_P2_SCA}
{\text{(P2-$i$)}}\quad \underset{\boldsymbol{\mathrm{w}}} {\max}\quad f(\mathbf{w}), \quad \mathrm{s.t.}\quad\text{(\ref{eqn_BeamFormer_Cons_P2})}, \text{(\ref{eqn_IntCons_P2})} \nonumber
\end{equation}
which is a quadratically constrained quadratic programming (QCQP) problem and thus can be optimally solved by the interior-point algorithm. The SCA algorithm proceeds to (P2-$(i+1)$) by setting $\mathbf{w}_{i+1}$ as the optimal solution to (P2-$i$), until convergence is reached. The computational complexity for solving each (P2-$i$) is given by $\mathcal{O}(N^3K^{1.5})$.

\subsection{Optimizing $\mathbf{T}$ for Given $\mathbf{w}$}
In this subproblem, our objective is to optimize the APV $\mathbf{T}$ with given ST's beamforming $\mathbf{w}$. Accordingly, (P1) can be simplified as
\begin{equation}\label{eqn_OptPrblm_P3}
{\text{(P3)}}\quad\underset{\mathbf{T}} {\max}\quad |\mathbf{h}_0^H(\mathbf{t})\mathbf{w}|^2, \quad \mathrm{s.t.}\quad \text{(\ref{eqn_Intf_Cons})}, \text{(\ref{eqn_MA_Region_Cons})}, \text{(\ref{eqn_MA_Coordinate_Cons})}\nonumber
\end{equation}
However, (P3) remains difficult to be optimally solved due to the intractability of the highly non-linear objective function in terms of the MA positions $\mathbf{t}_n$, $\forall n \in \mathcal{N}$. To circumvent this issue, we apply a similar discretization approach to \cite{W_Mei_Graph}.

Specifically, we uniformly sample the horizontal/vertical dimension of the transmit region ${\cal C}_t$ into $M$ ($M\gg N$) discrete points, with a spacing given by $\delta_s=A/M$. By this means, the continuous transmit region is discretized into $M^2$ sampling points, with the position of the $(i,j)$-th sampling point given by $\mathbf{p}_{i,j}=[-\frac{A}{2}+\frac{iA}{M},-\frac{A}{2}+\frac{jA}{M}]^T$, $i,j\in\mathcal{M}\triangleq\{1,2,\cdots,M\}$. Let $\mathcal{P}=\{\mathbf{p}_{i,j}|i,j\in\mathcal{M}\}$ denote the set of all sampling points. Next, we propose a sequential search method to optimize the MA's positions.

First, we construct a set of initial positions of MAs denoted by $\mathbf{\Tilde{t}}_n\in\mathcal{P}$, $\forall n \in \mathcal{N}$. We consider that the position of the $n$-th MA, i.e., $\mathbf{\Tilde{t}}_n$, needs to be updated in the $n$-th iteration of the sequential search, with the positions of all other $(N-1)$ MAs being fixed. Let $\mathcal{P}_n$ denote the set of all feasible sampling points for the $n$-th MA, which is given by
\begin{equation}\label{eqn_DiscreteSet}
    \mathcal{P}_n=\{\mathbf{p}\in\mathcal{P}|\,\,||\mathbf{p}-\mathbf{\Tilde{t}}_m||_2\ge D_{\min},\forall m \in\mathcal{N},m\ne n\}.
\end{equation}
Let $\mathbf{\Tilde{T}}_n=[\mathbf{\Tilde{t}}_1,\cdots,\mathbf{t}_n,\cdots,\mathbf{\Tilde{t}}_N]^T$ denote the APV of all $N$ MAs in the $n$-th iteration of the sequential search. Then, in this iteration, we optimize $\mathbf{t}_n$ by solving the following problem,
\begin{equation}\label{eqn_OptPrblm_P3}
{\text{(P3-$n$)}}\quad \underset{\mathbf{t}_n} {\max}\quad |\mathbf{h}_0^H(\mathbf{\Tilde{T}}_n)\mathbf{w}|^2, \quad \mathrm{s.t.}\quad \mathbf{t}_n \in\mathcal{P}_n,\,\, \text{(\ref{eqn_Intf_Cons})}\nonumber
\end{equation}
which can be optimally solved by performing an enumeration over $\mathcal{P}_n$. Denote by $\hat{\mathbf{t}}^*_n$ the optimal solution to (P3-$n$). Next, we update $\tilde{\mathbf{t}}_n=\hat{\mathbf{t}}^*_n$ and proceed to solving (P3-$(n+1)$). It can be seen that the computational complexity of the sequential search is given by $\mathcal{O}(NM^2)$.

    




        

Based on the above, we can alternately solve problems (P2-1) and (P3) by applying the SCA and the sequential search algorithms. As this process always yields a non-decreasing objective value of (P1), the convergence of the AO algorithm is always ensured.

\section{Numerical Results}
\begin{figure*}[!t]
\centering
\subfloat[]{
    \includegraphics[width=0.3\linewidth]{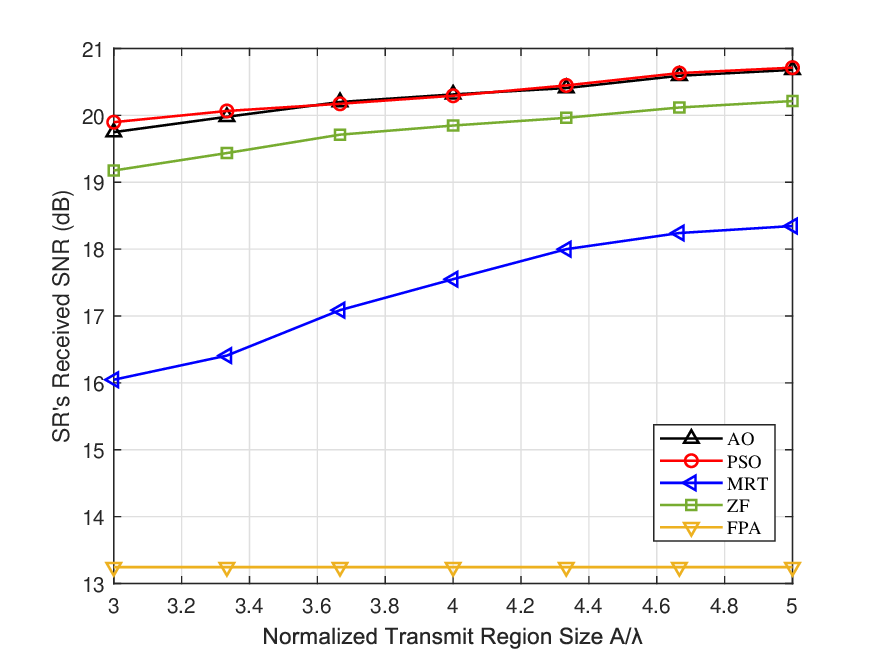}
}
\hfil
\subfloat[]{
    \includegraphics[width=0.3\linewidth]{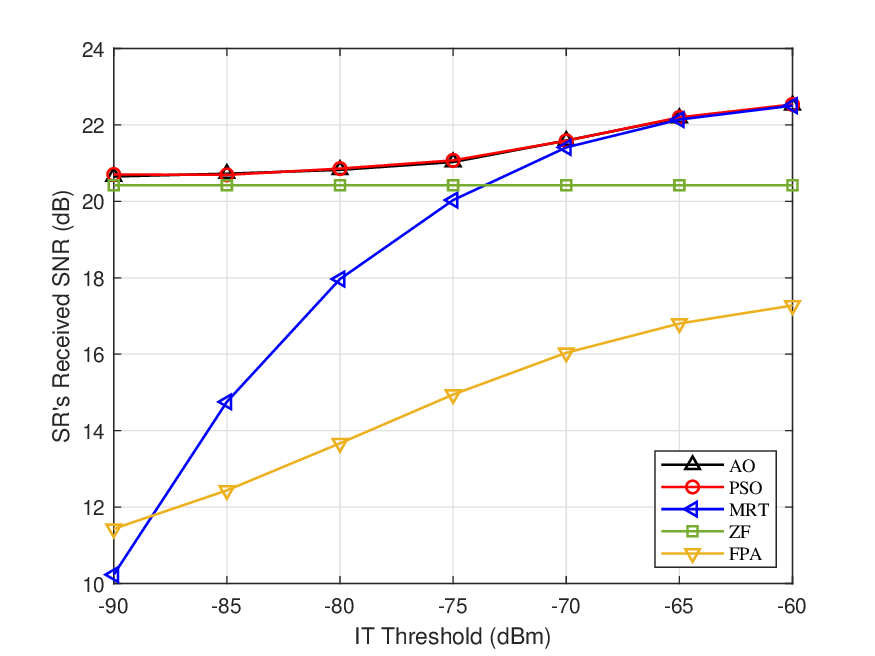}
}
\hfil
\subfloat[]{
    \includegraphics[width=0.3\linewidth]{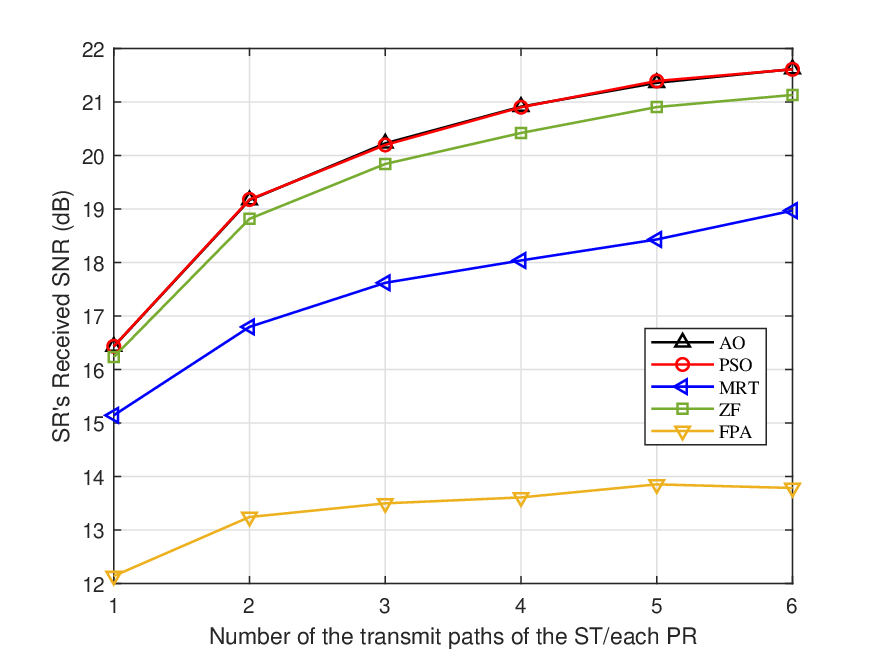}
}
\caption{SR's received SNR versus (a) the normalized transmit region $A/\lambda$; (b) the IT threshold; and (c) the number of transmit paths of the SR/each PR.}
\label{fig_sim}
\vspace{-19pt}
\end{figure*}

In this section, numerical results are provided to show the efficacy of our proposed algorithms. Unless otherwise specified, the simulation settings are as follows. The ST is equipped with $N=4$ MAs. The total number of PRs is set to $K=3$. The minimum distance between any two adjacent MAs is set as $D_{\min}=\frac{\lambda}{2}$ with $\lambda=0.1$ meter (m). The number of sampling points along $x$ and $y$ axes is set as $M=100$. The distance between the ST and the SR/PRs is assumed to be a random variable following uniform distribution between 20 m and 100 m. The numbers of transmit paths for the SR and each PR is assumed to be identical as $L=L_k=4, k \in {\cal K} \cup \{0\}$. The path gain $\beta_{k,p}$'s are assumed to follow the CSCG distribution, i.e., $\beta_{k,p}\sim\mathcal{CN}(0,\rho d_k^{-\alpha}/L),k \ic{K}\cup \{0\}, 1 \le p \le L$, where $\rho$ represents the path loss at the reference distance of 1 m and $\alpha=2.8$ denotes the path-loss exponent. The elevation and azimuth AoAs for each transmit path are assumed to be independent and identically distributed (i.i.d.) variables following the uniform distribution over $[-\pi /2, \pi /2]$. The ST's maximum transmit power $P_{\max}$ is $23$ dBm, and the average noise power $\sigma^2$ is $-80$ dBm. All the results are averaged over $100$ independent channel realizations.

Moreover, our proposed algorithm is compared with the following benchmark schemes:
\begin{enumerate}
    \item \textbf{Particle Swarm Optimization (PSO)}: The ST's transmit beamforming is optimized similarly to Section IV-A, while the positions of MAs are optimized in the continuous space via the PSO method, similarly to \cite{Z_Xiao_Multiuser}.
    \item \textbf{MRT}: The ST applies the MRT, and its MAs' positions are optimized via a similar sequential search as in Section IV-B.
    \item \textbf{Zero forcing (ZF)}: As $N\ge K$, the ST can apply ZF beamforming to null its interference to the $K$ PRs, i.e., $\mathbf{w}_{\mathrm{ZF}}(\mathbf{T})=\sqrt{P_{\max}}\frac{\mathbf{\hat{w}}(\mathbf{T})}{||\mathbf{\hat{w}}(\mathbf{T})||_2}$,
    where $\mathbf{\hat{w}}(\mathbf{T})=[\mathbf{I}_N-\mathbf{R}(\mathbf{T})(\mathbf{R}^H(\mathbf{T})\mathbf{R}(\mathbf{T}))^{-1}\mathbf{R}^H(\mathbf{T})]\mathbf{h}_0(\mathbf{T})$ and $\mathbf{R}(\mathbf{T})=[\mathbf{h}_1(\mathbf{T}),\mathbf{h}_2(\mathbf{T}),\cdots,\mathbf{h}_N(\mathbf{T})]$. Moreover, the MA positions are optimized via a similar sequential search as in Section IV-B.
    \item \textbf{FPA}: The ST's antennas are uniformly deployed in ${\cal C}_t$ with half-wavelength antenna spacing, and its transmit beamforming is optimized similarly to Section IV-A.
\end{enumerate}

First, Fig. 2(a) shows the SR's received SNR versus the normalized size of the transmit region, $A/\lambda$, by different schemes. The IT threshold is set to $\Gamma=-80$ dBm. It is observed that all schemes employing MAs outperform the FPA-based scheme, and their performance improves with increasing the transmit region size. This is because a larger transmit region endows the MAs with larger spatial degrees of freedoms to create more favorable channel conditions. It is also observed that our proposed AO algorithm can achieve a comparable performance to the PSO-based scheme with more efficient implementation. In addition, the performance gap between the MRT-based scheme and AO is observed to decrease with $A$. This implies that the interference mitigation capability of MRT becomes stronger for a larger size of antenna aperture, which is consistent with our analytical results provided in Section III.

Next, Fig. 2(b) shows the SR's received SNR versus the IT threshold, with $A=4\lambda$. It is noted that when $\Gamma$ is sufficiently large (e.g., $\Gamma\ge-70$ dBm), the MRT-based scheme achieves comparable performance to the AO and PSO, as expected. This is because for a sufficiently large $\Gamma$, the IT constraints can be approximately relaxed, and the ST can apply the MRT to maximize the received signal power at the SR as in the MRT-based scheme. In contrast, when the interference threshold $\Gamma$ is sufficiently small, the performance of the MRT-based scheme degrades, as the ST's transmit power may be limited in this case to satisfy the IT constraints. Instead, the ZF-based scheme is observed to achieve a close performance to AO and PSO.

Lastly, we plot the SR's received SNR versus the number of transmit paths of the SR/each PR (i.e., $L$) in Fig. 2(c), with $A=4\lambda$ and $\Gamma=-80$ dBm. It is observed that all schemes employing MAs outperform the FPA-based scheme and their performance gap increases with $L$. This is because more transmit paths lead to more significant small-scale fading, thus providing more significant spatial diversity for the MA position optimization. In addition, the proposed AO algorithm is observed to achieve close performance to PSO and significantly outperform other benchmark schemes.

\section{Conclusion}
In this letter, we studied the joint transmit beamforming and MA position optimization for a CR system, aiming to maximize the received signal power at the SR subject to the IT constraints for multiple PRs. Our theoretical analyses revealed that MAs can effectively mitigate the co-channel interference even with MRT, especially if the transmit region is sufficiently large. We also propose an AO algorithm to obtain high-quality suboptimal solutions to the MA position optimization problem. Numerical results showed the superiority of our proposed algorithms over the PSO algorithm and the conventional FPAs. It would be interesting to investigate low-latency implementation methods for MAs in future, e.g.,  electronically driven equivalent movement.

\end{document}